\documentclass{jaa}

\usepackage{graphicx}
\usepackage{amsmath}
\usepackage{xcolor}
\usepackage{natbib}
\usepackage{footmisc}
\usepackage{rotating,longtable}
\usepackage{lineno}
\usepackage{natbib}
\usepackage[T1]{fontenc}
\usepackage{ae,aecompl}

\begin{document}\sloppy

\title{LAMOST J010103.13+275449.6: A New Binary Exhibiting a Switch in Photometric Period from 1.7216\,d to 0.8444\,d}


\author{Y. H. Chen\textsuperscript{1*}}

\affilOne{\textsuperscript{1}Institute of Astrophysics, Chuxiong Normal College, Chuxiong 675000, China\\}


\twocolumn[{

\maketitle

\corres{yanhuichen1987@126.com}

\msinfo{ }{ }

\begin{abstract}
LAMOST J010103.13+275449.6 is bright with an apparent magnitude of 10.764 in the SDSS $g$ band. We searched for two low resolution spectra from LAMOST, apparent magnitudes from GALEX, Pan-STARRS, 2MASS, and WISE, X-ray data from XMM-Newton, and light curve from TESS. It is very interesting that the photometric period of light curve switch from 1.7216\,d to 0.8444\,d. In previous large sample statistical studies, it was identified as an eclipsing binary, a rotational variable star, or an emission line star. The ultraviolet excess in 2006 from GALEX, the hard X-ray emission in 2016 from XMM-Netwon, the high amplitude photometric variability in 2019 from TESS, together with the nearly featureless photospheric spectra in 2011 and 2015 from LAMOST, collectively reduce the likelihood of the rapidly rotating single star magnetic activity scenario. The spectral fitting for the two main-sequence components shows that the system is a main-sequence binary consisting of F and G type stars. Mathematical modeling suggests that, with a 1.7216\,d orbital period and tidal locking of the companion, the period switching of the light curve can be semi-quantitatively explained by orbital motion, involving a large spot at the substellar point in the early phase and small spots at both the substellar and antipodal points in the later phase. The binary parameters are preliminary with large uncertainties and more in-depth studies can be carried out in the future based on more spectra that may be released by LAMOST.
\end{abstract}

\keywords{binary:individual(LAMOST J010103.13+275449.6); magnetic activity; close binary}

}]


\doinum{12.3456/s78910-011-012-3}
\artcitid{\#\#\#\#}
\volnum{000}
\year{0000}
\pgrange{1--}
\setcounter{page}{1}
\lp{1}


\section{Introduction}

Stars are among the most common luminous objects in the universe, and a substantial fraction of them are found in binary systems. Winters et al. (2019) studied 1120 M dwarf primaries within 25\,pc via trigonometric parallaxes and reported that the stellar multiplicity rate is 26.8 $\pm$ 1.4\% and the stellar companions rate is 32.4 $\pm$ 1.4\%. Raghavan et al. (2010) conducted an in-depth study of 454 F6-K3 type stars within 25\,pc from the Hipparcos catalog and found that the proportion of binaries is at least 33 $\pm$ 2\%. About 90\% of all stars fall on the main sequence (MS) and about 98\% (Winget \& Kepler 2008) of all stars will evolve to be white dwarfs (WDs). In 2018, El-badry \& Rix constructed wide binaries within 200\,pc from Gaia DR2 data, including more than 50,000 MS/MS binaries, 3,000 MS/WD binaries, and 400 WD/WD binaries. Studying binaries is of universal significance.

Binaries contain a wealth of physical laws. The previously described categories such as double MS stars, MS/WD binaries, and double WDs are all classifications based on the evolutionary status of the component stars. According to different observational methods, binaries can be classified into visual binaries, spectroscopic binaries, eclipsing binaries, and astrometric binaries. According to the different shapes of their light curves, eclipsing binaries can be classified into EA Algol type, EB $\beta$ Lyrae type, and EW W Ursae Maijoris type (Heintz 1978). Combined orbital solution of eclipsing binaries (photometric plus spectroscopic) can yield complete physical parameters of the two component stars, including their masses, radii, orbital inclination, effective temperature ($T_{\rm eff}$) ratio, distance, and other related quantities. According to the extent to which the component stars fill their Roche lobes, binaries can be classified into detached binaries, semi-detached binaries, contact binaries, and overcontact binaries. Cataclysmic variables (CVs) are semi-detached binary systems consisting of a WD and a MS (or a giant), in which mass transfer occurs, making them important laboratories for studying mass transfer. Ultra-short orbital period double WDs are important sources for gravitational wave detection (Burdge et al. 2019).

Stellar magnetic activity, encompassing processes such as dynamo action, magnetic reconnection, starspot evolution, and flaring, remains a fundamental open question at the forefront of modern stellar physics. Strassmeier (2009) reviewed the physical nature of starspots, namely that magnetic fields suppress convective motion, and carried out simulation studies of starspots on various types of MS stars. Mathur et al. (2014) investigated 22 Kepler F-type stars with rotation periods less than 12\,d and found a variety of magnetic activity behaviours, ranging from cyclic modulation and long-lived active regions to monotonic trends and no apparent cycles. Balona (2013) analysed two years of Kepler photometry of A-type stars and suggested that the light variations in about 40\% arise from rotational modulation by starspots or other corotating structures, while flares are detected in about 1.5\%. Stellar magnetic activity is a very common phenomenon.

Research on stellar magnetic activity in binary systems is also one of the hot topics in binary star physics. Messina (2008) analyzed about 38,000 high-precision photoelectric nightly observations in the U, B, and V filters, conducting a thorough investigation of 14 famous magnetically active close binary stars. Yu et al. (2025) investigated the chromospheric activity of double MS stars and revealed that tidal forces significantly influence surface activity. K\H{o}v$\acute{a}$ri (2025) provided a brief review of the dynamo mechanisms operating in RS CVn and BY Dra type binaries, focusing on spot cycles, active longitudes, flare activity, and differential rotation. While studying the stellar spectra released by the Large Sky Area Multi-Object Fiber Spectroscopic Telescope (LAMOST), we serendipitously discovered LAMOST J010103.13+275449.6. We found that the source, based on its approximately 20-day high-precision light curve from Transiting Exoplanet Survey Satellite (TESS, Ricker et al. 2014), exhibits a phenomenon where the photometric period switches from 1.7216\,d to 0.8444\,d. This source should be a very interesting target for studying stellar magnetic activity, whether it is a single or a binary star. In Sect. 2, we analyze the LAMOST low resolution search (LRS) spectrum, multi-band SED spectra, and TESS light curve of LAMOST J010103.13+275449.6. The double MS spectral fitting for LAMOST J010103.13+275449.6 are performed by Binary Tools (Liu et al. 2024) in Sect. 3. In Sect. 4, we conduct a preliminary light curve fitting for LAMOST J010103.13+275449.6 based on a python code for geometric starspot modeling. At last, we give a discussion and conclusions in Sect. 5.

\section{LAMOST LRS spectrum, multi-band SED spectra, and TESS light curve of LAMOST J010103.13+275449.6}

The Chinese national scientific research facility LAMOST has a 20 square degree field of view, a large effective aperture ranging from 3.6 to 4.9 meters in diameter, and can obtain 4,000 spectra in a single exposure (Zhao et al. 2012). From October 24, 2011, to June 30, 2025, the LAMOST DR13 dataset has officially released over 30.82 million spectra, providing an extremely rich spectral data resource for various astrophysical research. Cui et al. (2012) reported that the LAMOST LRS spectra cover a wavelength range of 3,700-9,000\,{\AA} with a spectra resolution of 1,800 (Stoughton et al. 2002, Abazajian et al. 2003) at 5,500\,{\AA}, including a blue channel optimized for 3,700-5,900\,{\AA} and a red channel for 5,700-9,000\,{\AA}.

\begin{figure}
\begin{center}
\includegraphics[width=8.5cm,angle=0]{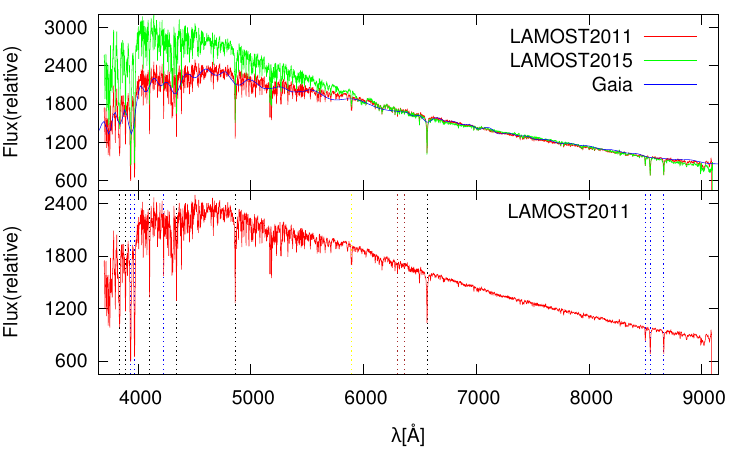}
\end{center}
\caption{Spectra figure of LAMOST J010103.13+275449.6, including two LAMOST LRS spectra and one Gaia XP spectrum. The vertical dashed lines in different colors indicate different absorption lines.}
\end{figure}

\begin{table*}
\begin{center}
\scriptsize
\caption{Table of SNRs and atmospheric parameters for two LAMOST LRS spectra of LAMOST J010103.13+275449.6. The SNR is the signal to noise ratio and the second column values are SNR of $u$, $g$, $r$, $i$, and $z$ bands respectively.}
\begin{tabular}{llllllllllll}
\hline
MJD             &SNR of $u$,$g$,$r$,$i$,$z$                    &$T_{\rm eff}$[K]      &log\,$g$           &[Fe/H]             &RV[km/s]           \\
\hline
55898.518103125 &\,50.01,\,159.34,\,254.96,\,251.75,\,187.62   &6235.71 $\pm$ 21.61   &4.424 $\pm$ 0.028  &0.018 $\pm$ 0.015  &-0.5 $\pm$ 3.66    \\
57326.605462500 &\,141.14,\,359.98,\,539.66,\,580.76,\,387.58  &6193.06 $\pm$ 15.76   &4.381 $\pm$ 0.019  &0.012 $\pm$ 0.009  &-4.34 $\pm$ 2.87   \\
\hline
\end{tabular}
\end{center}
\end{table*}

\begin{table}
\scriptsize
\begin{center}
\caption{Table of wavelength of absorption lines identified in LAMOST LRS spectrum of LAMOST J010103.13+275449.6. They correspond by color to the vertical dashed lines in the lower panel of Fig. 1. Both the two spectra in Table 1 have these absorption lines.}
\begin{tabular}{llllllllllll}
\hline
Absorption lines               &Wavelength({\AA})                &Color        \\
\hline
H$\alpha$                      &\,6564.61                        &black        \\
H$\beta$                       &\,4862.68                        &black        \\
H$\gamma$                      &\,4341.68                        &black        \\
H$\delta$                      &\,4102.89                        &black        \\
H$\epsilon$                    &\,3971.19                        &black        \\
H$\zeta$                       &\,3890.15                        &black        \\
H$\eta$                        &\,3836.47                        &black        \\
Ca II                          &\,8500.35,\,8544.44,\,8664.52    &blue         \\
Ca I                           &\,4227.92                        &blue         \\
Ca II H                        &\,3969.59                        &blue         \\
Ca II K                        &\,3934.78                        &blue         \\
O I                            &\,6365.54                        &brown        \\
O I                            &\,6302.05                        &brown        \\
Na I                           &\,5895.6                         &yellow       \\
\hline
\end{tabular}
\end{center}
\end{table}

While studying the stellar spectra from the LAMOST LRS DR13, we serendipitously noticed LAMOST J010103.13+275449.6 (LAMOST J010103.01+275449.6). It was observed twice by LAMOST, on December 3, 2011 and October 31, 2015, as shown in Fig. 1. The signal to noise ratios (SNRs) and atmospheric parameters are shown in Table 1. The second column of Table 1 gives the SNRs of $u$, $g$, $r$, $i$, and $z$ bands respectively. Using the central wavelength and bandwidth of SDSS photometric system, each S/N value is taken as the median value of the pixels in that band. The atmospheric parameters are derived from the LAMOST Stellar Parameter Pipeline (LASP, Wu et al. 2014; Luo et al. 2015) directly. It is an F8 or F7 type MS star labeled in the LAMOST data release. Judging from the errors, the values of $T_{\rm eff}$ and surface gravity (log\,$g$) are reliable results from the single star fitting. Its metallicity is very close to the solar metallicity.

In Fig. 1, the prominent identified absorption lines are marked with vertical dashed lines in different colors. The elements and wavelengths of the absorption lines corresponding to the different colors are presented in Table 2. The Balmer absorption lines of hydrogen are very prominent, as indicated by the black dashed lines in Fig 1. The width of the Balmer lines suggests that the star is a MS star rather than a compact WD. These spectral lines are absorption lines, with no emission features even at the line cores. Two LAMOST spectra and one Gaia spectrum were compared after applying a scaling factor. Interestingly, the blue end flux of the LAMOST spectrum observed in 2015 is markedly higher than that of the other two spectra. It appears that there may be a minor issue with the flux calibration at the blue end of the 2015 spectrum. If it is not a flux calibration issue, it could very likely be magnetic activity of a single star, or magnetic activity of a component star in a binary system. Using local continuum normalization, the equivalent widths (EWs) of the first three Balmer lines (H$\alpha$, H$\beta$, and H$\gamma$) are 3.248 $\pm$ 0.430, 2.776 $\pm$ 0.464, and 2.413 $\pm$ 0.445 {\AA} for the 2011 spectrum, and 3.288 $\pm$ 0.427, 2.682 $\pm$ 0.441, and 2.380 $\pm$ 0.442 {\AA} for the 2015 spectrum. The differences in EWs are all within the uncertainties, and further detailed spectroscopic analysis would require higher spectral resolution.

\begin{table*}
\tiny
\begin{center}
\caption{The apparent magnitude table for LAMOST J010103.13+275449.6. The observations are from GALEX, Pan-STARRS, SDSS, 2MASS, and WISE respectively. Two SDSS observations were conducted in 2008 and 2009, respectively.}
\begin{tabular}{llllllllllllllll}
\hline
               &\multicolumn{2}{c}{GALEX}&\multicolumn{5}{c}{Pan-STARRS}                                   &\multicolumn{3}{c}{2MASS}          &\multicolumn{4}{c}{WISE}                    \\
\hline
filter         &FUV          &NUV        &$g$          &$r$        &$i$         &$z$          &$y$         &$J$         &$H$        &$Ks$      &$W1$      &$W2$      &$W3$       &$W4$      \\
$\lambda$[$nm$]&\,154.885    &\,230.337  &\,481.016    &\,615.547  &\,750.303   &\,866.836    &\,961.360   &\,1,235     &\,1,662    &\,2,159   &\,3352.6  &\,4602.8  &\,11560.8  &\,22088.3 \\
\hline
mag            &\,20.3971    &\,14.8537  &\,10.654     &\,10.352   &\,10.239    &\,10.189     &\,10.175    &\,9.304     &\,9.086    &\,9.012   &\,8.968   &\,8.998   &\,8.960    &\,8.540   \\
err            &\,0.2217     &\,0.0107   &             &           &\,0.001     &             &            &\,0.020     &\,0.022    &\,0.020   &\,0.021   &\,0.020   &\,0.025    &          \\
\hline
               &             &           &\multicolumn{5}{c}{SDSS}                                         &            &           &          &          &          &           &          \\
\hline
filter         &             &           &$u$          &$g$        &$r$         &$i$          &$z$         &            &           &          &          &          &           &          \\
$\lambda$[$nm$]&             &           &\,360.804    &\,467.178  &\,614.112   &\,745.789    &\,892.278   &            &           &          &          &          &           &          \\
\hline
mag(2008)      &             &           &\,12.270     &\,10.764   &\,10.249    &\,10.168     &\,10.892    &            &           &          &          &          &           &          \\
err(2008)      &             &           &\,0.001      &\,0.001    &\,0.001     &\,0.001      &\,0.002     &            &           &          &          &          &           &          \\
\hline
mag(2009)      &             &           &\,14.088     &\,10.781   &\,10.258    &\,10.174     &\,12.756    &            &           &          &          &          &           &          \\
err(2009)      &             &           &\,0.005      &\,0.001    &\,0.001     &\,0.001      &\,0.010     &            &           &          &          &          &           &          \\
\hline
\end{tabular}
\end{center}
\end{table*}

We search the data from the VizieR online data catalog and display the apparent magnitude and corresponding errors for LAMOST J010103.13+275449.6 in Table 3. The apparent magnitude data are sourced from Bianchi et al. (2017) for GALEX, Chambers et al. (2016) for Pan-STARRS, Ahn et al. (2012) for SDSS, Cutri et al. (2003) for 2MASS, and Cutri et al. (2012) for WISE, respectively. The center wavelength of these filters are from the SVO Filter Profile Service. This is a bright target source, with an apparent magnitude of 10.764 in the SDSS $g$ band.  However, it is strange that the apparent magnitudes in the u band and z band differ by about 2 magnitudes in the two SDSS observations from 2008 and 2009, respectively. No obvious infrared excess phenomenon is detected. However, it appears to show an ultraviolet excess on November 7, 2006 (Bai et al. 2018). The SDSS has not measured a spectrum for LAMOST J010103.13+275449.6 and marks it as a galaxy or a star. Gaia measured a distance of 159.32 $\pm$ 0.41\,pc for this star and rule out the probability of being a galaxy.

\begin{figure}
\begin{center}
\includegraphics[width=8.5cm,angle=0]{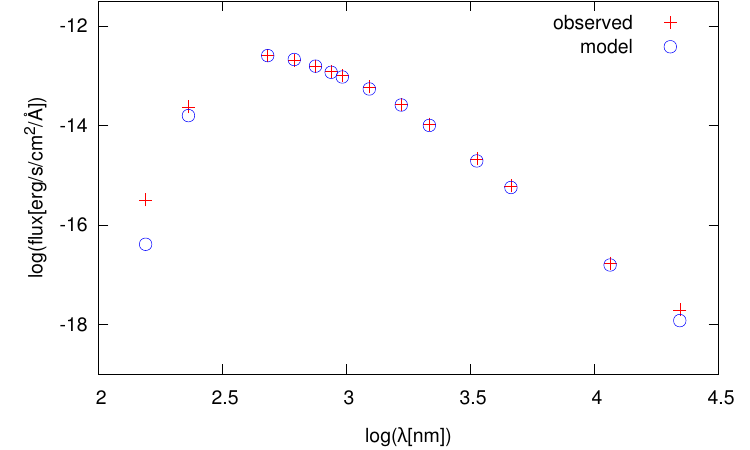}
\end{center}
\caption{VOSA SED fitting of LAMOST J010103.13+275449.6. The observed data are from Table 3.}
\end{figure}

\begin{table}
\scriptsize
\begin{center}
\caption{Parameters of the best fitting model in Fig. 2.}
\begin{tabular}{llllllllllll}
\hline
                        &$T_{\rm eff}$[K]   &log\,$g$         &M[$M_{\odot}$] &R[$R_{\odot}$]  \\
\hline
Parameters              &\,6000$\pm$50      &\,4.5$\pm$0.25   &1.36$\pm$0.78  &1.09$\pm$0.02   \\
\hline
\end{tabular}
\end{center}
\end{table}

The virtual observatory Spectral Energy Distribution (SED) analyzer (VOSA, Bayo et al. 2008) is a convenient user-friendly web-based SED fitting tool. Because the zero-magnitude fluxes differ across different bands, SED fitting is very helpful for identifying infrared excess and ultraviolet excess. Adopting the ATLAS9 Kurucz ODFNEW/NOVER models (Castelli \& Kurucz 2003) and BT-Settl models (Allard et al. 2012), we performed SED fitting for LAMOST J010103.13+275449.6 on the first row of apparent magnitude data from Table 3, as shown in Fig. 2. This object is relatively nearby, with extinction $A_{v}$ $\approx$ 0. The assessment of the infrared excess is affected by the uncertainties in the W4 data, while the ultraviolet excess is indeed obvious. This could be due to magnetic activity of a single star or originate from a close binary. The best-fit model comes from the BT-Settl models, and its parameters, presented in Table 4, are generally consistent with those in Table 1.

\begin{figure}
\begin{center}
\includegraphics[width=8.5cm,angle=0]{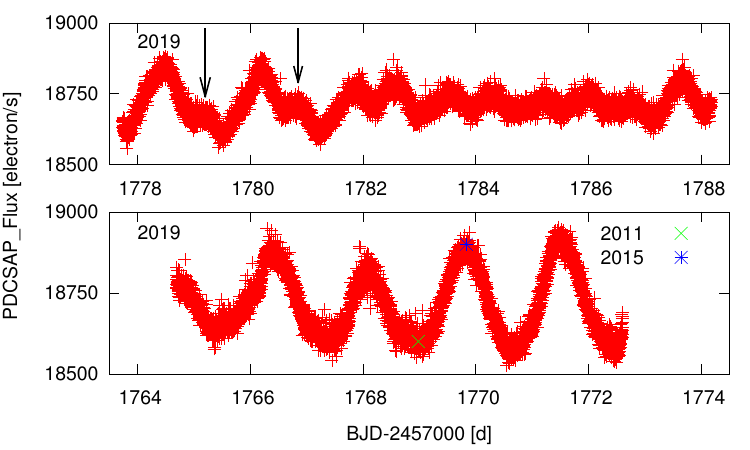}
\end{center}
\caption{TESS light curve figure of LAMOST J010103.13+275449.6, which was observed from October 8, 2019 to November 2, 2019. The green cross and blue asterisk represent the time of the LAMOST LRS spectrum in 2011 (BJD = 2455899.0180346\,d) and 2015 (BJD = 2457327.1054408\,d), respectively, converted using a period of 1.7216\,d.}
\end{figure}

The study of light curves can provide more valuable information. The Barbara A. Mikulski Archive for Space Telescopes (MAST, Hou et al. 2023) provides a rich archive of astronomical observations from missions like Hubble, Kepler, TESS, and so on. From October 8, 2019, TESS observed this source (TIC 15535353) for nearly 25 days, with exposure times of 120\,s (provenance name of SPOC) and 1800\,s (provenance name of QLP and TESS-SPOC). The physical principles reflected by different exposure times are consistent. For the 120\,s exposure data, we removed low quality data points and used PDCSAP\_Flux to create Fig. 3. The 120\,s data originally contained 18,012 data points, which decreased to 13,024 after quality cuts. Using the frequency extraction software Period04 (Lenz \& Breger 2005), we obtained a photometric variability period of 1.7216 $\pm$ 0.0011\,d for the light curve in the lower panel of Fig. 3, and a period of 0.8444 $\pm$ 0.0010\,d for the light curve in the upper panel starting from BJD - 2457000 = 1781.91\,d. Clearly, the photometric period has switched from 1.7216\, to 0.8444\,d. Moreover, the black arrow in the upper panel indicates that this is a gradual process. Through phase-folding fitting, we obtained the following formula describing the times of minima.
\begin{equation}
T_{1}=1765.5088 \pm 0.0021 + (1.7216 \pm 0.0011) \times E_{1}
\end{equation}
\begin{equation}
T_{2}=1782.1766 \pm 0.0031 + (0.8444 \pm 0.0010) \times E_{2}
\end{equation}
\noindent The Eq\,(1) is for the lower panel of Fig. 3 and the Eq\,(2) is for the upper panel of Fig. 3. From the lower panel of Fig. 3, the photometric variability amplitude reaches a level of $\sim$2\%. The specific MJD values of the two LAMOST LRS spectra were converted to BJD and scaled using a period of 1.7216\,d yielding the epochs of the green cross (2011) and the blue asterisk (2015), as displayed in Fig. 3. The green cross and the blue asterisk were then artificially shifted along the ordinate to align with the locations of the TESS observational data points. Interestingly, the converted 2011 LAMOST spectrum is located near the flux minimum, whereas the converted 2015 LAMOST spectrum is located near the flux maximum. This might be related to the disagreement with the spectral flux distribution presented in Fig. 1. In their study of stellar parameters and magnetic activity properties of TESS objects observed in the LAMOST survey, Zhang et al. (2023) classified LAMOST J010103.13+275449.6 as exhibiting stellar activity, derived a period of 1.7543\,d, and interpreted it as a rotational variable star. The spectral type (F7) and atmospheric parameters derived by Zhang et al. (2023) are consistent or close to those released by LAMOST. However, statistical studies with large samples can easily identify longer periods and miss halving periods. Gao et al. (2025) also reported a rotational period of 1.7543\,d for LAMOST J010103.13+275449.6. However, both Green et al. (2023) and Ijspeert et al. (2004) identified it as an eclipsing binary. In addition, Webb et al. (2020) reported this source as an X-ray source in the XMM-Newton serendipitous source catalogue. In the XMM-Newton observations from June 28 to 29, 2016, the ep\_8\_flux is 1.29 $\pm$ 0.106 $\times$ $10^{-13}$ erg/$cm^{2}$/s, the hardness ratio HR1 is 0.66 $\pm$ 0.02, and the probability of being AGN is 55.6\%. The distance provided by Gaia rules out the possibility of it being an extragalactic galaxy or an AGN.

We first ruled out single-star pulsations. Handler (2013) conducted a review of all pulsating variables and reported that only a small fraction exhibit pulsation periods on the order of days. Pulsators with periods on the order of or close to one day are usually found to be multi-periodic (Zhou 2026), such as $\delta$ Scuti, $\gamma$ Doradus, Solar-like oscillators, and so on. Few monoperiodic $\delta$ Scuti (HADS, Zhou \& Liu 2003) stars show light curve amplitudes substantially larger than the amplitude displayed in Fig. 3. Then, we examine the magnetic activity of a rapidly rotating F8-type MS star. Among the 33 F-type MS stars with abundant X-ray emission reported by Shimura et al. (2025), HD 193825, the one with the strongest X-ray radiation, exhibits an HR1 value of 0.58 $\pm$ 0.03, which is smaller than that of our target source (0.66 $\pm$ 0.02). Metanomski et al. (1998) studied the F, G, and K stars observed by ROSAT all-sky survey and reported a mean value of HR1 to be 0.13 $\pm$ 0.35. According to Savanov (2019), the light curve amplitude of our source is at the upper boundary of the amplitude distribution for F-type stars. Interpreting the light curve as magnetic activity on a single star would require invoking the flip-flop phenomenon (Elstner \& Korhonen 2005) on a rapidly rotating F8 star, which is rare. The multi-epoch observations from GALEX (2006), XMM-Newton (2016), and TESS (2019) consistently demand strong magnetic activity. Conversely, the LAMOST spectra from 2011 and 2015, together with the Gaia XP spectra, show no obvious anomalies in their photospheric features. This phenomenon substantially undermines the single star hypothesis. 

We prefer that LAMOST J010103.13+275449.6 is a binary system with an orbital period of 1.7216\,d. Both the LAMOST LRS spectra and the multi-band SED spectra can rule out the possibility of a compact object companion. A double MS star system is the optimal solution. The light curve is inconsistent with the simple eclipse pattern of an eclipsing binary. The light curve should be the overall manifestations of physical processes such as shallow eclipses in the binary system, magnetic activity of the component star, ellipsoidal effect, and reflection effect. Magnetic activity occurring at the starspot and antipodal points could potentially give rise to period switching in the light curve. The ultraviolet excess can be attributed to the binary system, while the relatively hard X-ray emission can be accounted for by the coronae of the component stars.

\section{Double MS spectral fitting for LAMOST J010103.13+275449.6 by Binary Tools}

\begin{figure*}
\begin{center}
\includegraphics[width=12.8cm,angle=0]{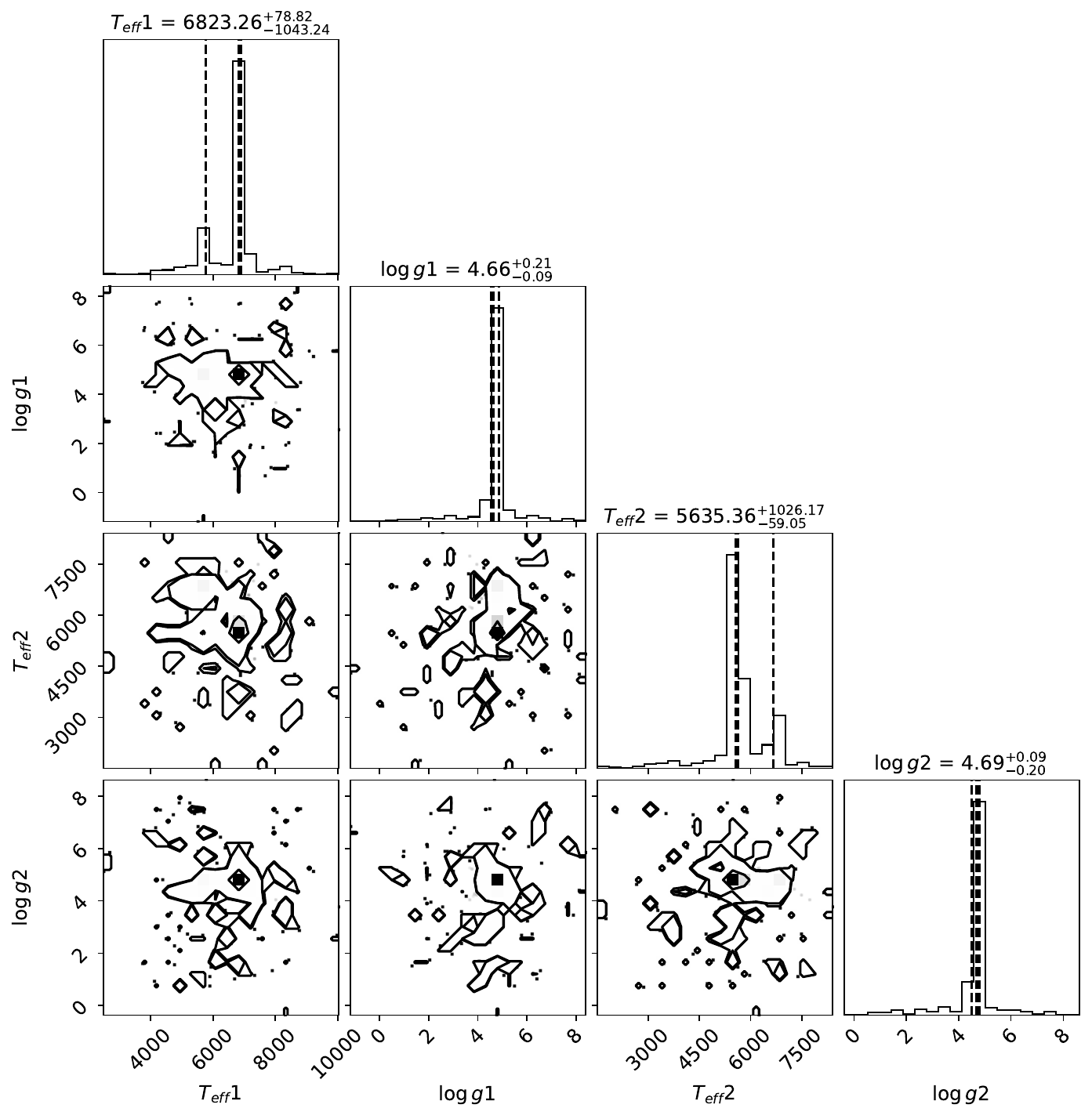}
\end{center}
\caption{Corner figure for fitting LAMOST J010103.13+275449.6 spectrum by Binary Tools. The contours show the joint confidence regions and correlations between two parameters, while the vertical dashed lines mark the median and the 1$\sigma$-equivalent uncertainties for each individual parameter.}
\end{figure*}

\begin{figure}
\begin{center}
\includegraphics[width=8.5cm,angle=0]{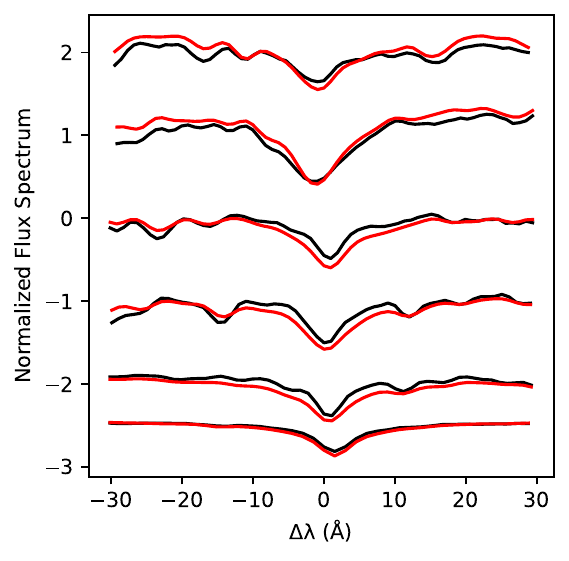}
\end{center}
\caption{Balmer absorption line fitting figure. From bottom to top, the Balmer lines are H$\alpha$, H$\beta$, H$\gamma$, H$\delta$, H$\epsilon$, and H$\zeta$ respectively.}
\end{figure}

Liu et al. (2024) developed a spectral decomposition code, Binary Tools, that is designed for low resolution spectra and is capable of simultaneously fitting both components in close binary systems. We use Binary Tools to perform double MS star spectral fitting on the second LAMOST LRS spectrum, which was observed in 2015 and has higher SNRs in Table 1. The MS spectra employed by Binary Tools are drawn from the Coelho synthetic stellar library (Coelho 2014), a high resolution theoretical database covering B to M type stars across 2,500-9,000 {\AA} with $T_{\rm eff}$ = 3,000-25,000 K. Binary Tools uses an artificial neural network to generate MS spectra, which significantly enhances both the efficiency and accuracy of the spectral synthesis.

\begin{figure*}
\begin{center}
\includegraphics[width=13.8cm,angle=0]{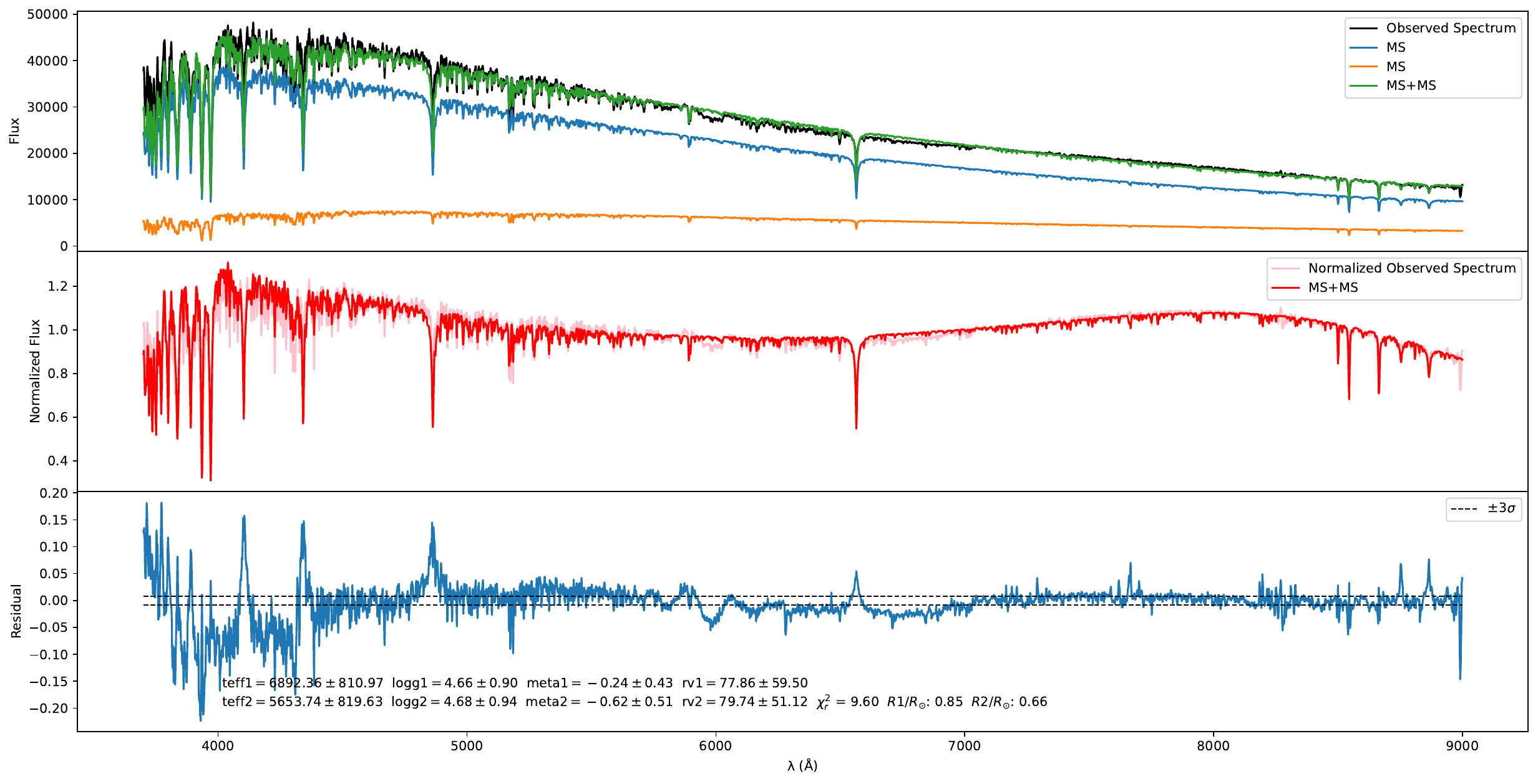}
\end{center}
\caption{Figure of blackbody fitting validation for LAMOST LRS spectrum and Gaia XP spectrum. Spectral fits figure of the two MS stars, along with the normalized spectra and residuals.}
\end{figure*}

We initially attempted input with a higher $T_{\rm eff}$ (6500\,K) star and a lower $T_{\rm eff}$ (5500\,K) star, and the final fitting corner figure, Balmer absorption line profiles figure, and double MS spectral figure are presented in Fig. 4, Fig. 5, and Fig. 6 respectively. Although the corner plot indicates relatively large errors in the effective temperatures of the two components, the fitting does yield a convergent solution. The Balmer lines from H$\alpha$ to H$\zeta$ are well fitted, as shown in Fig. 5. Figure 6 shows the detailed fits of the two model spectra to the observed spectrum. We tried multiple sets of input parameters and obtained consistent fitting results in all cases. It should be noted that the neural network in Binary Tools maps stellar parameters to rest-frame spectra and is consequently insensitive to RV. In addition, the low spectral resolution of LAMOST further restricts the precision of RV measurements. Together, these limitations render RV estimates unreliable in the fitting of low resolution spectra of close binary systems. In fact, the RVs of the two spectra in Table 1 are -0.5 $\pm$ 3.66 km/s and -4.34 $\pm$ 2.87 km/s, respectively. The low  values are consistent with the binary configurations (near the substellar and antipodal points) during the two epochs of observation shown in Fig. 3. In addition, the Gaia RVS observations span 13 visibility periods over a time duration of 920.85\,d, yielding a RV of 0.06 $\pm$ 0.49 km/s with an amplitude of 7.48 km/s. This suggests that the binary system may have a small intrinsic RV or a small orbital inclination. Actually, the observed RV is that of the composite spectrum, which is affected by many factors such as the rotation of individual stars and the orbital motion of the binary system. It is very difficult to inversely derive the RVs of the two component stars from the existing observational data.

\begin{table*}
\begin{center}
\scriptsize
\caption{Table of spectral fitting results by Binary Tools.}
\begin{tabular}{lllllllllllll}
\hline
           &$T_{\rm eff}$\,1 &log\,$g$\,1 &[Fe/H]\,1 &M\,1           &R\,1           &$T_{\rm eff}$\,2 &log\,$g$\,2  &[Fe/H]2  &M\,2           &R\,2           \\
           &[K]              &            &          &[$M_{\odot}$]  &[$R_{\odot}$]  &[K]              &             &         &[$M_{\odot}$]  &[$R_{\odot}$]  \\
\hline
           &                 &            &          &               &\multicolumn{3}{c}{Binary Tools}               &         &               &               \\
\hline
value      &6892             &4.66        &-0.24     &1.20           &0.85           &5654             &4.68         &-0.62    &0.77           &0.66           \\
err$\pm$   &811              &0.90        &0.43      &0.52           &0.90           &820              &0.94         &0.51     &0.95           &0.83           \\
err+       &79               &0.21        &0.18      &0.01           &0.05           &1026             &0.09         &0.43     &0.35           &0.23           \\
err-       &1043             &0.09        &0.22      &0.46           &0.26           &59               &0.20         &0.05     &0.06           &0.09           \\
\hline
\end{tabular}
\end{center}
\end{table*}

In Table 5, we present a detailed parameter table from the Binary Tools fitting, consistent with Fig. 4, 5, and 6. The values and errors (err$\pm$) in Table 5 are derived from the parameters corresponding to the Maximum A Posteriori (MAP), while the values and errors (err+ and err- in Table 5) in Fig. 4 are derived from the median and the 68\% confidence interval. Decomposing the spectra of binary stars using low-resolution spectroscopy is inherently very difficult, especially when the observed spectra may also contain signatures of strong magnetic activity. The $T_{\rm eff}$ and surface gravity values are reliable in terms of order of magnitude, and this binary is an F-type plus G-type binary system. The mass ratio of the binary system is of reference value. In addition, for the LAMOST LRS spectrum observed in 2011, we did not obtain a satisfactory fitting result, possibly owing to its relatively low SNRs.

\section{Light curve fitting for LAMOST J010103.13+275449.6 based on a python code for geometric starspot modeling}

In Sect. 4, we plan to study the modeling of the light curve of LAMOST J010103.13+275449.6. The light curve amplitude in the lower panel of Fig. 3 reaches $\sim$2\%. We analyze the reflection effect in the binary first. Using the parameters of Table 5, we estimate the upper limit of the reflection amplitude ($A_{max}$) as the ratio of the hot star's luminosity intercepted by the cool star's cross sectional area to the total luminosity (Napier 1971), as shown below,
\begin{equation}
A_{\text{max}} = \frac{(4\pi R_{\text{hot}}^2 \sigma T_{\text{hot}}^4) \cdot \frac{\pi R_{\text{cool}}^2}{4\pi d^2}}{4\pi R_{\text{hot}}^2 \sigma T_{\text{hot}}^4 + 4\pi R_{\text{cool}}^2 \sigma T_{\text{cool}}^4}.
\end{equation}
\noindent This yields a light curve amplitude upper limit of 0.15\%, which fails to explain the observations. In practice, the reflection effect is considerably more complicated than the simple geometric approximation presented in Eq.\,(3). When accounting for the albedo of the cooler component, the resulting amplitude is expected to be further reduced.

The observed light curves might originate from a variety of complex physical mechanisms, such as tidal locking, starspots induced by magnetic activity, ellipsoidal variations, the reflection effect, and limb darkening. These processes present significant challenges to model fitting. To address this, we employ a purely empirical mathematical model to approximate the primary physical process, magnetic starspot activity. The light curves are first normalized by their median values and then phase-folded based on Eq.\,(1) and (2). We assume that the angular radii of the spots at the substellar and antipodal points are $\theta_1$ and $\theta_2$ respectively. The projected area of the spot at rotational phase $\phi$ is modeled as,
\begin{equation}
f_{cool}(\phi)=\frac{1-\cos(\theta)}{2} \times \sin(i) \times \cos[(2\pi(\phi-\phi_0))].
\end{equation}
\noindent In Eq.\,(4), i is the binary orbital inclination (also axial tilt of the companion star under tidal locking) and $\phi_0$ is the shifted reference zero epoch to that in Eq.\,(1) and Eq.\,(2).
The corresponding model flux of the cool star and the binary is then given by,
\begin{equation}
F_{cool}(\phi)=1+[(\frac{T_{spot}}{T_{cool}})^{4}-1] \times f_{cool}(\phi),
\end{equation}
\begin{equation}
F_{total}(\phi)=1+(\frac{T_{cool}}{T_{hot}})^{4} \times (\frac{R_{cool}}{R_{hot}})^{2} \times [F_{cool}(\phi)-1].
\end{equation}
\noindent When the starspot is facing the Earth, the flux reaches its minimum. The maximum of the light curve modulation is likely to be closely related to the limb darkening effect. We therefore account for both effects together through the approximate treatment described by the above formulas. Using the parameter ranges summarized in Table 6, we perform separate fits to the two observational data sets shown in Fig. 3. In the first data set, we assume a single large starspot at the substellar point. In the second data set, we assume the presence of two smaller spots of identical size, one at the substellar point and the other at the antipodal point (Mathematically, this is achieved by halving the period and taking the equivalent of one sunspot). A total of 67,500 models were used to fit the observational data in Fig. 3. The best-fit models corresponding to the three sets of orbital inclinations are listed in Table 6, with the fitting results for the first parameter set displayed in Fig. 7.

\begin{table*}
\begin{center}
\scriptsize
\caption{BTable of starspot parameters and binary orbital inclination for LAMOST J010103.13+275449.6.}
\begin{tabular}{llllllllllll}
\hline
            &$P_{\text{pho}}$ &$T_{spot1}$[K]&$\theta_1$[degree]&$T_{spot2}$[K]  &$\theta_2$[degree] &$\phi_0$               &$i$[degree]     &$\chi^{2}$  \\
\hline
range,step  &                 &4000-4800,200 &20-38,2           &4000-5000,200   &10-28,2             &(-0.01)-(-0.002),0.002 &25-65,5         &            \\
\hline
            &1.7216\,d        &4800          &30                &                &                   &-0.008                 &$50$             &0.031540    \\
            &0.8444\,d        &              &                  &4800            &14                 &-0.004                 &$50$             &0.011508    \\
\hline
            &1.7216\,d        &4600          &32                &                &                   &-0.008                 &$35$             &0.031543    \\
            &0.8444\,d        &              &                  &5000            &18                 &-0.004                 &$35$             &0.011508    \\
\hline
            &1.7216\,d        &4000          &30                &                &                   &-0.008                 &$30$             &0.031592    \\
            &0.8444\,d        &              &                  &4000            &14                 &-0.004                 &$30$             &0.011507    \\
\hline
\end{tabular}
\end{center}
\end{table*}

\begin{figure*}
\begin{center}
\includegraphics[width=12.8cm,angle=0]{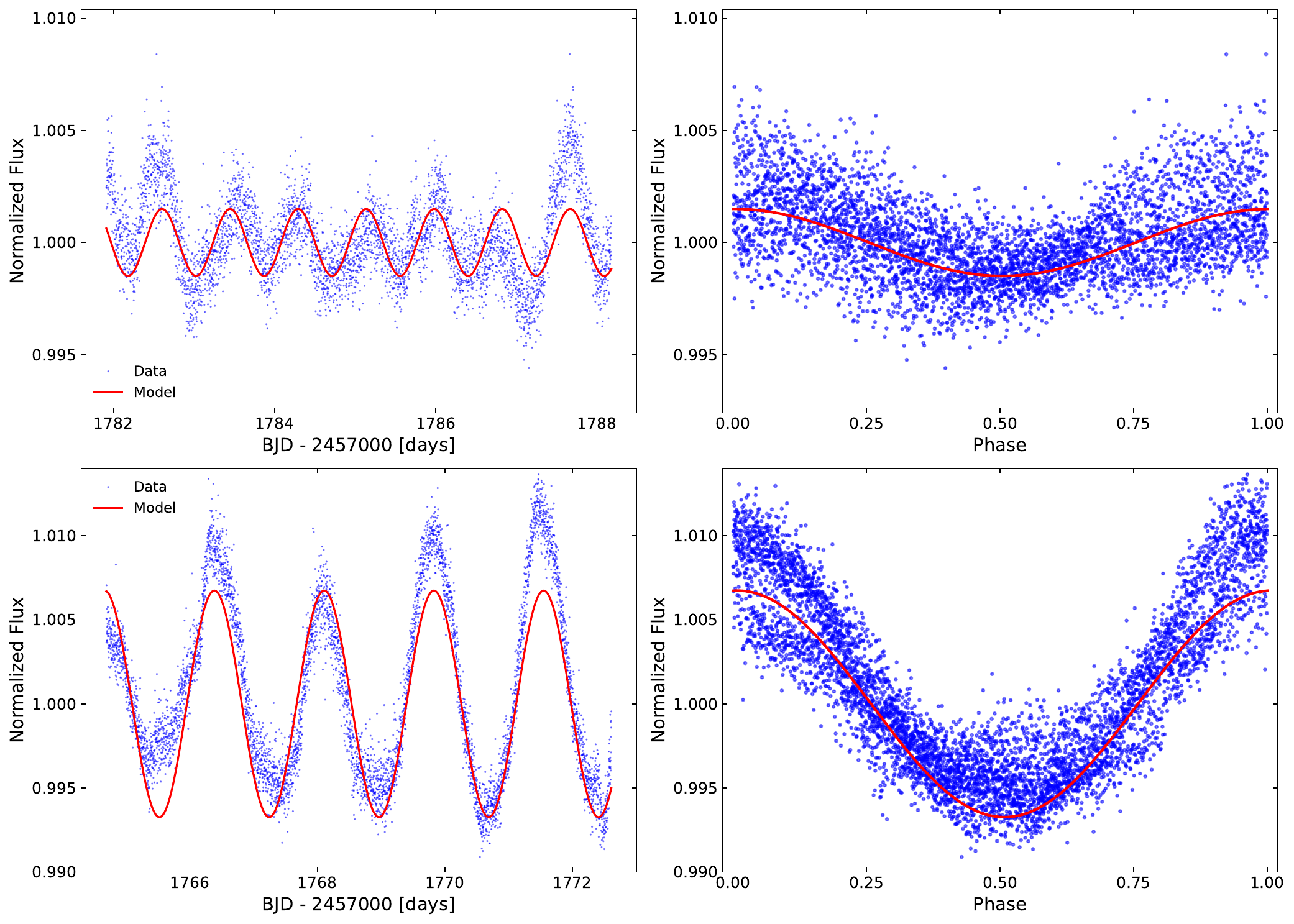}
\end{center}
\caption{Light curve fit figure based on the best-fit model, shifting 0.5 phase to T0 in Eq.\,(1) and (2).}
\end{figure*}

As can be seen from Table 6, a low binary orbital inclination can be compensated by increasing the spot size or decreasing the spot temperature to achieve a larger modulation amplitude. The spot angular radii and the temperature contrasts relative to the companion's effective temperature ($T_{\rm eff}$\,2 = 5654\,K), as listed in Table 6, are all within physically reasonable ranges. Using high-resolution spectroscopy of HR 1099, Strassmeier \& Bartus (2000) reported a spot angular radius of 30 degrees and a temperature contrast of 1100\,K. Following the method described in Jie et al. (2025), we used the two LAMOST spectra displayed in Fig. 1 to calculate the ratio of the chromospheric Ca II H \& K line fluxes to the bolometric flux, commonly referred to as the log$R'_{HK}$ index. Note that the LAMOST spectral line wavelengths are given in vacuum. It is -4.1 for the two LAMOST spectra in Fig. 1, indicating strong chromospheric magnetic activity, or even saturation. The strong chromospheric magnetic activity is in agreement with the ultraviolet excess and the relatively hard X-ray emission discussed in Sect. 2. As shown in Fig. 7, the mathematical model captures only the dominant component of the light curves, leaving the fine structures around the maximum poorly fitted. This indicates the intrinsic complexity of the physical processes governing the binary system. The reduction in modulation amplitude from the first data set to the second may also indicate the evolution or transport of magnetic activity on the stellar surface.

\section{A Discussion and Conclusions}

When studying the released LRS spectra from LAMOST, we accidentally noticed LAMOST J010103.13+275449.6 (or LAMOST J010103.01+275449.6). It is bright, with an apparent magnitude of 10.764 in the SDSS $g$ band. In the open source spectral data, we only found two LAMOST spectra and one Gaia XP spectrum. The value of log$R'_{HK}$ (Jie et al. 2025) is -4.1 for both LAMOST spectra, demonstrating strong chromospheric magnetic activity. The SED fitting of the multi-band apparent magnitudes from GALEX, Pan-STARRS, 2MASS, and WISE indicates that this source exhibits an ultraviolet excess, as shown in Fig. 2. In addition, in the XMM-Newton observations, both the higher ep\_8\_flux (1.29 $\pm$ 0.106 $\times$ $10^{-13}$ erg/$cm^{2}$/s) and the larger HR1 (0.66 $\pm$ 0.02) indicate the presence of relatively hard X-ray radiation. What is most interesting is that we have obtained the TESS light curve data, which captures the physical process of the photometric variability period switching from 1.7216\,d to 0.8444\,d, as shown in Fig. 3.

In the literature, LAMOST J010103.13+275449.6 has been identified as an eclipsing binary (Green et al. 2023, IJspeert et al. 2024), a rotational variable star (Gao et al. 2023), or an emission-line star(Zhang et al. 2023). The SDSS DR16 marked it as a star or a galaxy and the XMM-Newton marked it as an AGN. The distance (159.32 $\pm$ 0.41\,pc) from Gaia directly exclude the possibility of this source being a galaxy or an AGN. The ultraviolet excess in 2006 from GALEX, the hard X-ray emission in 2016 from XMM-Netwon, the high amplitude photometric variability in 2019 from TESS, together with the nearly featureless photospheric spectra in 2011 and 2015 from LAMOST, collectively reduce the likelihood of the rapidly rotating single star magnetic activity scenario. A close binary with a 1.7216\,d orbital period offers a reasonable explanation that matches most of the observations. 

It is intrinsically difficult to disentangle the spectra of two stars with similar parameters from low resolution spectra. Using the Binary Tools (Liu et al. 2024) code, we performed repeated calculations and obtained a set of converged solutions, as shown in Fig. 6. It is most likely an F+G type binary star. It is worth noting that our measured RVs are inaccurate. As is evident from the line cores in Fig. 5, deriving RVs of a binary from low resolution spectra is quite challenging. The Gaia RVS observations, covering 13 visibility periods, provide a Rv of 0.06 $\pm$ 0.49 km/s with an amplitude of 7.48 km/s, which is indicative of a low orbital inclination. We adopt a scenario in which the companion star is tidally locked, with a large starspot located at the substellar point during the early phase, and two small starspots of equal size located at the substellar point and its antipodal point during the late phase. This can explain the period transition observed in the light curve in Fig. 3. The latter two solutions with small inclinations in Table 6 can produce smaller RVs.

Binary systems contain rich physical processes. Our analysis of the binary nature of this source is robust, while the derivation of binary parameters remains preliminary and subject to large uncertainties. In the future, if LAMOST releases more low resolution and medium resolution spectra of LAMOST J010103.13+275449.6, we will perform more in-depth studies.

\section*{Acknowledgments}

This work made use of the data from LAMOST (Large Sky Area Multi-Object Fiber Spectroscopic Telescope, also known as the Guoshoujing Telescope) (https://cstr.cn/31118.02.LAMOST). LAMOST is a Chinese national mega-science facility, operated by National Astronomical Observatories, Chinese Academy of Sciences. The authors acknowledge the support of the Physics Discipline Construction Project of Chuxiong Normal College (Grant No. 20260600016). This publication makes use of VOSA, developed under the Spanish Virtual Observatory (https://svo.cab.inta-csic.es) project funded by MCIN/AEI/10.13039/501100011033 through grant PID2023-146210NB-I00. VOSA has been partially updated by using funding from the European Union's Horizon 2020 Research and Innovation Programme, under Grant Agreement $n^\circ$ 776403 (EXOPLANETS-A).






\begin{thebibliography}{99}
\bibitem[\protect\citeauthoryear{Abazajian et al}{2003}]{Abazajian2003} Abazajian K., Adelman-McCarthy J. K., Ag$\ddot{u}$eros M. A., et al., 2003, AJ, 126, 2081
\bibitem[\protect\citeauthoryear{Ahn et al}{2012}]{Ahn2012} Ahn C. P., Alexandroff R., Prieto C. A., et al., 2012, ApJS, 203, 21
\bibitem[\protect\citeauthoryear{Allard et al}{2012}]{Allard2012} Allard B. F., Homeier D., Freytag B., 2012, Phil. Trans. R. Soc. A, 370, 2765
\bibitem[\protect\citeauthoryear{Bai et al}{2018}]{Bai2018} Bai Y., Liu J. F., Wicker J., et al., 2018, ApJS, 235, 16
\bibitem[\protect\citeauthoryear{Balona et al}{2013}]{Balona2013} Balona L. A., 2013, MNRAS, 431, 2240
\bibitem[\protect\citeauthoryear{Bayo et al}{2008}]{Bayo2008} Bayo A., Rodrigo C., Barrado y Navascu$\acute{e}$s D., et al., 2008, A\&A, 492, 277
\bibitem[\protect\citeauthoryear{Bianchi et al}{2017}]{Bianchi2017} Bianchi L., Shiao B., Thilker D., 2017, ApJS, 230, 24
\bibitem[\protect\citeauthoryear{Burdge et al}{2019}]{Burdge2019} Burdge K. B., Coughlin M. W., Fuller J., et al., 2019, Nature, 571, 528
\bibitem[\protect\citeauthoryear{Castelli et al}{2003}]{Castelli2003} Castelli F. \& Kurucz R. L., 2003, IAUS, 210, A20
\bibitem[\protect\citeauthoryear{Coelho}{2014}]{Coelho2014} Coelho P. R. T., 2014, MNRAS, 440, 1027
\bibitem[\protect\citeauthoryear{Chambers et al}{2016}]{Chambers2016} Chambers K. C., Magnier E. A., Metcalfe N., et al., 2016, arXiv:1612.05560, DOI:10.48550/arXiv.1612.05560
\bibitem[\protect\citeauthoryear{Cui et al}{2012}]{Cui2012} Cui X. Q., Zhao Y. H., Chu Y. Q., et al., 2012, RAA, 12, 1197
\bibitem[\protect\citeauthoryear{Cutri et al}{2003}]{Cutri2003} Cutri R. M., Skrutskie M. F., van Dyk S., et al., 2003, yCat, 2246, 0
\bibitem[\protect\citeauthoryear{Cutri et al}{2012}]{Cutri2012} Cutri R. M., et al., 2012, yCat, 2311, 0
\bibitem[\protect\citeauthoryear{Elstner et al}{2005}]{Elstner2005} Elstner D., \& Korhonen H., 2005, Astronomische Nachrichten, 326, 278
\bibitem[\protect\citeauthoryear{El-Badry et al}{2018}]{El-Badry2018} El-Badry K., \& Rix H. W., 2018, MNRAS, 480, 4884
\bibitem[\protect\citeauthoryear{Gao et al}{2025}]{Gao2025} Gao X. Y., Chen X. D., Wang S., et al., 2025, ApJS, 276, 57
\bibitem[\protect\citeauthoryear{Green et al}{2023}]{Green2023} Green M. J., Maoz D., Mazeh T., et al., 2023, MNRAS, 522, 29
\bibitem[\protect\citeauthoryear{Handler}{2013}]{Handler2013} Handler G., 2013, Asteroseismology, Planets, Stars and Stellar Systems Vol. 4. Springer, Berlin, P. 207 (ISBN 978-94-007-5614-4)
\bibitem[\protect\citeauthoryear{Heintz et al}{1978}]{Heintz1978} Heintz W. D. 1978, Double Stars, Vol. 15 (D. Reidel Publishing Company)
\bibitem[\protect\citeauthoryear{Hou et al}{2023}]{Hou2023} Hou W., Luo A. L., Dong Y. Q., et al., 2023, ApJ, 165, 148
\bibitem[\protect\citeauthoryear{Ijspeert et al}{2024}]{Ijspeert2024} Ijspeert L. W., Tkachenko A., Johnston C., et al., 2024, A\&A, 691, A242
\bibitem[\protect\citeauthoryear{Jie et al}{2025}]{Jie2025} Jie Y., Charlotte G., Saskia H., et al., 2025, NatAs, 9, 1045Y
\bibitem[\protect\citeauthoryear{Kovari}{2025}]{Kovari2025} K\H{o}v$\acute{a}$ri Zs., 2025, CoSka, 55, 182
\bibitem[\protect\citeauthoryear{Lenz et al}{2005}]{Lenz2005} Lenz P., \& Breger M., 2005, Commun. Asteroseismol., 146, 53
\bibitem[\protect\citeauthoryear{Liu et al}{2024}]{Liu2024} Liu G. H., Tang B. T., Ren L. L., et al., 2024, A\&A, 690, A29
\bibitem[\protect\citeauthoryear{Luo et al}{2015}]{Luo2015} Luo A. L., Zhao Y. H., Zhao G., et al., 2015, RAA, 15, 1095
\bibitem[\protect\citeauthoryear{Mathur et al}{2014}]{Mathur2014} Mathur S., Garc$\acute{i}$a R. A., Ballot J., et al., 2014, A\&A, 562, A124
\bibitem[\protect\citeauthoryear{Messina}{2008}]{Messina2008} Messina S., 2008, A\&A, 480, 495
\bibitem[\protect\citeauthoryear{Metanomski}{1998}]{Metanomski1998} Metanomski A. D. F., Pasquini L., Krautter J., et al., 1998, Astron. Astrophys. Suppl. Ser., 131, 197
\bibitem[\protect\citeauthoryear{Napier et al}{1971}]{Napier1971} Napier W. Mcd., 1971, Ap\&SS, 11, 475
\bibitem[\protect\citeauthoryear{Raghavan et al}{2010}]{Raghavan2010} Raghavan D., Mcalister H. A., Henry T. J., et al., 2010, ApJS, 190, 1
\bibitem[\protect\citeauthoryear{Ricker et al}{2014}]{Ricker2014} Ricker G. R., Winn J. N., Vanderspek R., et al., 2014, SPIE, 9143, 20
\bibitem[\protect\citeauthoryear{Savanov}{2019}]{Savanov2019} Savanov I. S., 2019, Astrophysical Bulletin, 74, 431
\bibitem[\protect\citeauthoryear{Shimura et al}{2025}]{Shimura2025} Shimura T., Mitsuishi I., Kunitomo M., et al., 2025, arXiv:2509.18677, DOI:10.48550/arXiv.2509.18677
\bibitem[\protect\citeauthoryear{Stoughton et al}{2002}]{Stoughton2002} Stoughton C., Lupton R. H., Bernardi M., et al., 2002, AJ, 123, 485
\bibitem[\protect\citeauthoryear{Strassmeier et al}{2000}]{Strassmeier2000} Strassmeier K. G., \& Bartus J., 2000, A\&A, 354, 537
\bibitem[\protect\citeauthoryear{Strassmeier et al}{2009}]{Strassmeier2009} Strassmeier K. G., 2009, A\&AR, 17, 251
\bibitem[\protect\citeauthoryear{Webb et al}{2020}]{Webb2020} Webb N. A., Coriat M., Traulsen I., et al., 2020, A\&A, 641, A136
\bibitem[\protect\citeauthoryear{Winget et al}{2008}]{Winget2008} Winget D. E., \& Kepler S. O., 2008, ARA\&A, 46, 157
\bibitem[\protect\citeauthoryear{Winters et al}{2019}]{Winters2019} Winters J. G., Henry T. J., Jao W. C., et al., 2019, AJ, 157, 216
\bibitem[\protect\citeauthoryear{Wu et al}{2014}]{Wu2014} Wu Y., Du B., Luo A. L., et al., 2014, IAUS, 306, 340
\bibitem[\protect\citeauthoryear{Yu et al}{2025}]{Yu2025} Yu J., Gean C., Hekker S., et al., 2025, Nature Astronomy, 9, 1045
\bibitem[\protect\citeauthoryear{Zhang et al}{2023}]{Zhang2023} Zhang L. Y., Su T. H., Misra P., et al., 2023, ApJS, 264, 17
\bibitem[\protect\citeauthoryear{Zhao et al}{2012}]{Zhao2012} Zhao G., Zhao Y. H., Chu Y. Q., et al., 2012, RAA, 12, 723
\bibitem[\protect\citeauthoryear{Zhou et al}{2026}]{Zhou2026} Zhou A. Y., 2026, NewA, 125, 102519
\bibitem[\protect\citeauthoryear{Zhou et al}{2003}]{Zhou2003} Zhou A. Y., \& Liu Z. L., 2003, AJ, 126, 2462
\end{thebibliography}
\end{document}